\documentclass[fleqn,12pt, twoside]{article}
\usepackage[headings]{espcrc2}
\usepackage[figuresright]{rotating}
\usepackage{graphicx}
\usepackage{amssymb}

\newcommand{\AmS}{{\protect\the\textfont2
  A\kern-.1667em\lower.5ex\hbox{M}\kern-.125emS}}

\title{Radiation efficiencies of the pulsars detected in the optical range}
\author{Zharikov S.\address[1]{OAN IA UNAM, Ensenada, BC, Mexico}, 
Shibanov Yu.\address[2]{Ioffe Physical Technical Inst, RAS, St. Petersburg, 
Russia}, Komarova V.\address[3]{Special Astrophysical Observatory, RAS, Russia}}
\runtitle{Radiation efficiencies of the pulsars.}
\runauthor{S. Zharikov, Yu. Shibanov, V. Komarova}
\begin{document}
\begin{abstract}
Using available multiwavelength data for the pulsars of different ages
detected in the optical range,  we analyze the efficiencies of the
conversion of the pulsar spindown power ${\rm \dot{E}}$ into the observed
non-thermal luminosity ${\rm L}$ in different spectral domains. We find
that these pulsars show significantly non-monotonic dependencies
of the optical and X-ray  efficiencies (${\rm\eta = L/\dot{E} }$) versus
pulsar age with a pronounced  minimum at the beginning of the middle-age
epoch (${\rm \sim 10^4}$~yr) and  comparably higher  efficiencies of  younger  and older pulsars.
In addition, we find a strong  correlation between the optical
and 2--10 keV X-ray luminosities of these pulsars that implies
the same origin of their nonthermal  emission in both spectral domains.
%\begin{keyword}
%Neutron stars, Pulsars
%\end{keyword}
\end{abstract}
%\begin{keyword}
%Neutron stars, Pulsars
%\end{keyword}
\maketitle
%\begin{document}
%%%%%%%%%%%%%%%%%%%%%%%%%%%%%
\section{Introduction}
%%%%%%%%%%%%%%%%%%%%%%%%%%%%%
\label{}
  The properties of the observed pulsar emission show that radio and
gamma-ray  photons   have  nonthermal  origin  and   are  produced  by
relativistic particles in different parts of the pulsar magnetosphere.
The soft  X-ray-optical radiation can be a combination of
several  nonthermal  spectral components  from  the magnetosphere  and
thermal components from the whole surface of a cooling neutron star (NS) 
and from its hot polar caps heated by the relativistic particles from the magnetosphere. 
 The nonthermal component has been found to dominate the optical emission of several
young, middle-aged, and old pulsars.  It is believed that nonthermal
components are powered by NS rotational energy loss $\dot{E}=4\pi^2 I
\dot{P}/{P}^3$, where 
%%% !
$I\approx 1.1\times 10^{45}\ gm\  cm^{2}$
%%% !
 is the inertia momentum
of a NS of 1.4 $M_\odot$ and 10~km radius, $P$ and $\dot{P}$ are the
pulsar period and its derivative, respectively. $\dot{E}$ is
frequently called also the spindown luminosity and denoted as
$L_{sd}$. The ratio $\eta =L/\dot{E}$, where $L$ is the photon
luminosity, is used to characterize the efficiency of the conversion
of $\dot{E}\equiv L_{sd}$ into the pulsar emission at a given spectral
domain.  In this paper, we analyze the evolution of the nonthermal
luminosities and efficiencies of pulsars detected in the optical
range. 
%%% !
Such analysis has been performed about 10~yr ago 
in pioneering works  \cite{Mereg,Goldoni}.    
However,  
considerable progress in the pulsar distance measurements, 
and much higher quality of recent optical 
and X-ray observations allowing to resolve better the nonthermal 
and thermal emission components from the pulsars and to identify  
in these ranges several new  objects enable us to update significantly  
previous analysis and to get qualitatively new results on 
the evolution of old pulsars.    
%%% !
%High accuracy of recent data 
%%% !
%on 
%the pulsar 
%%% !
%distances, 
%better resolution of the nonthermal and thermal emission components, 
%and a new data on old pulsars enables us to revise the results 
%of pioneering works on
% the multiwavelength phenomenology of pulsars \cite{Mereg,Goldoni}, 
% and to get qualitatively new results on evolution of old pulsars. 
%\vspace{-0.4cm}
\begin{table*}[t]
\caption{The dynamical ages ${\rm \tau =P/2\dot{P}}$, distances ${\rm d}$, spindown luminosities 
${\rm \dot{E}~(or~L_{sd})}$, 
and observed nonthermal luminosities in the radio, ${\rm L_{R}}$, optical, ${\rm L_{Opt}}$,
X-rays, ${\rm L_{X}}$, 
and gamma-rays, ${\rm L_{\gamma}}$, 
%ranges 
of seven radio pulsars detected in the 
optical range \cite{Zhar1,ZharS}. Figures in brackets are  
 ${\rm \pm1\sigma}$  uncertainties of the values. }
\begin{tabular}{llllllll}
\hline\hline
%           &		      &         	 &		       &			    &			       &		    &	  \\ 
Source     &log ${\tau}$ & d                &  log ${\rm \dot{E}}$&log ${\rm L_{R}}^a$	& log ${\rm L_{Opt}}$ &log ${\rm L_{X}}$ & log ${\rm L_{\gamma}}$\\ \vspace{-0.3cm}
           &                  &                  &                    &                  &            &           &         \\ 
           &[yr]              &  [pc]            & [{\rm erg~s$^{-1}$}]& $\rm [ mJy~kpc^2]$      & [{\rm erg~s$^{-1}$}]  &  [{\rm erg~s$^{-1}$}] &  [{\rm erg~s$^{-1}$}] \\ \vspace{-0.3cm}
           &                  &                  &                     &                  &            &           &         \\ 
           &                  &                  &         &  408 MHz  &    B band  &    2-10 keV        &  $ \ge$400MeV           \\ \hline          \vspace{-0.4cm}          
           &                                    &         &                           &                   &                  &               &  \\% \vspace{-0.5cm}
Crab       &  3.1	      &2.0$\times10^3$&  38.65  & 3.41(4)	  &33.23(5)		  &$36.67(^{+20}_{-26})$  & $35.7(^{+1}_{-3})$   \\  \vspace{-0.3cm}
           &	              &		         &	   &		&	  \\ 
B0540-69   & 3.2     &5.0$\times10^4$& 38.17   & 3.30(9)&33.47(15)		   &$36.99(^{+19}_{-23})$  &$ \le 35.97$ \\ \vspace{-0.3cm}
           &		      &	      	         &	   &		&	  \\ 
Vela       & 4.1    & ${293(^{+19}_{-17})} $   & 36.84   & 2.64(20)   &28.3(3)	   & $31.2(^{+36}_{-38})^i$ & $33.9(^{+1}_{-3})$  \\ \vspace{-0.3cm}
           &   		      &		         &	   &		&	  \\ 
B0656+14   &  5.0   & ${288(^{+33}_{-27})}$    & 34.58   & -0.27(9)	&27.53(8)		    & $30.30(^{+36}_{-28})$ & $32.37(^{+10}_{-30})$ \\ \vspace{-0.3cm}
           &		      &		         & 	   &	        &	 \\ 
Geminga    & 5.5      &${153(^{+59}_{-34})}$  & 34.51   &${0.375(^{+27}_{-23})}$&  ${26.95(^{+16}_{-10})}$& $29.35(^{+38}_{-36})$ &$32.95(^{+10}_{-30})$ \\ \vspace{-0.3cm}
           &		      &		         & 	   &	        &	 \\ 
B1929+10   & 6.5	  & 361$^{10}_{-8}$  & 33.59   & 1.52(5)    &${27.26(^{+20}_{-33})}$&$29.86(^{+13}_{-15})$ & $\le 32.57$	\\  \vspace{-0.3cm}
           &		      &		         & 	   &	        &	 \\ 
B0950+08   &  7.2	      &262(5)       & 32.75   & 1.44(16)	&26.88(8)	&$29.28(^{+13}_{-18})$ & $ \le 32.51$ \\  \hline
%           &		      &		         & 	   &	        &	&  \\ \hline

% \hline \\
\end{tabular}
\begin{tabular}{l}
{\footnotesize$^a$ The radio  luminosity is defined traditionally as  ${\rm L_{_{R}} =S_{408}d^2_1~[mJy~kpc^2]}$,
 where ${\rm S_{408}}$ is the observed 
 flux } \\ {\footnotesize density from  a pulsar   at 408 MHz in ${\rm mJy}$, and ${\rm d_1}$ is its distance in kpc. At a   
typical radio band  FWHM } \\ {\footnotesize
of  about 100 kHz  the conversion factor  
to  standard luminosity units  
is ${\rm \approx9.51\times10^{21}~[erg~s^{-1}]/[mJy~kpc^2]}$.}
% Errors in this }\\ {\footnotesize
% table and error bars in figures correspond to ${\rm \pm1\sigma}$  uncertainties of the values.}
\end{tabular}
\label{Pulsars}
%\vspace{-0.4cm}
\end{table*}  
\begin{figure}[t]
\setlength{\unitlength}{1mm}
\resizebox{8.7cm}{!}{\begin{picture}(80,115)(0,0)
\put ( 3,18) {\includegraphics[width=56mm, clip]{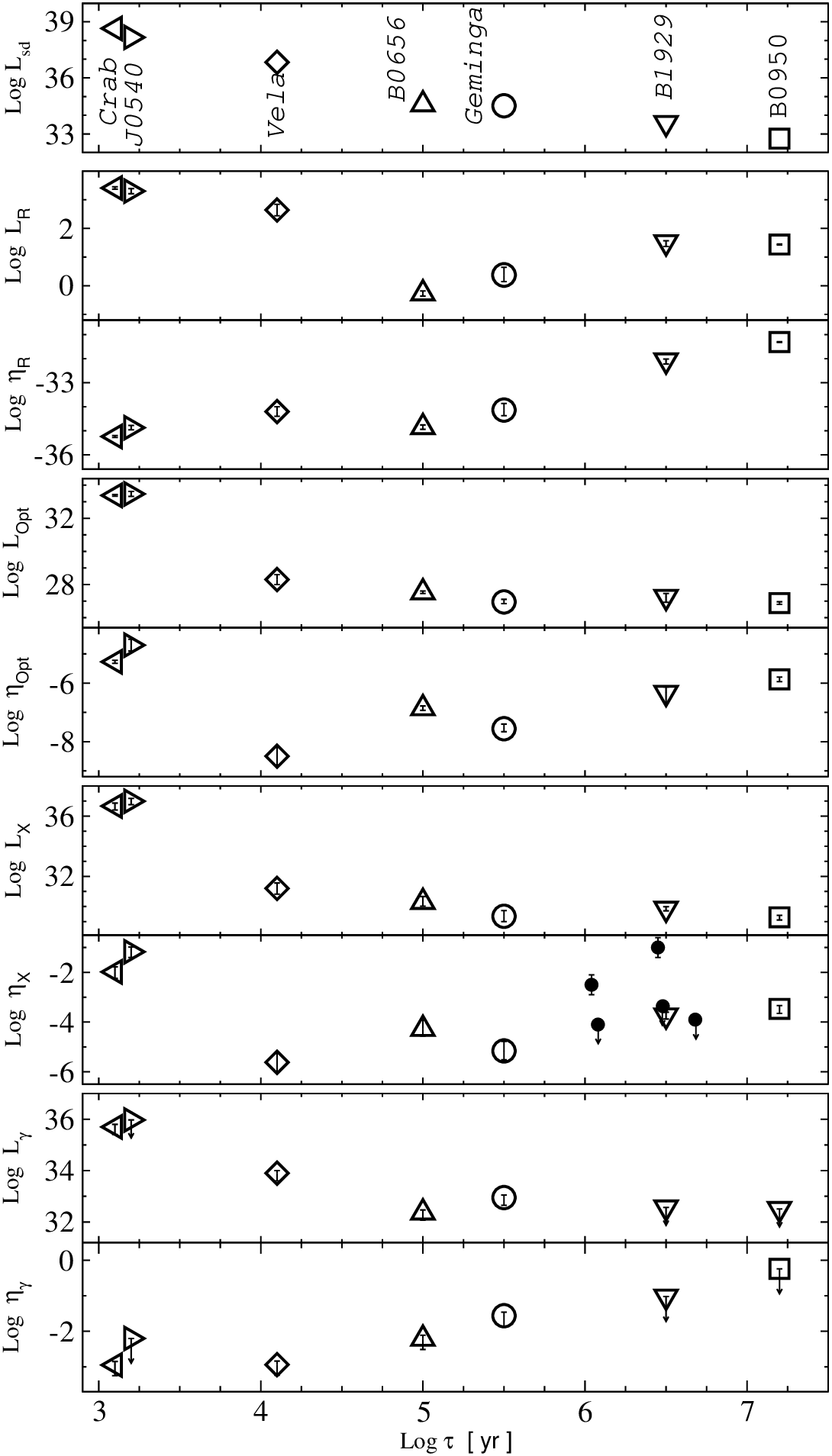}}
\put (-3, 27){ \parbox[t]{75mm} 
{ \caption{{\it From top to bottom:} evolution of the  pulsar spindown, 
radio, optical, X-ray, and gamma-ray  luminosities 
and respective efficiencies, with  dynamical age. 
Filled circles  are the data  taken  from \cite{Zav}.
Hereafter errorbars and arrows show ${\rm\pm1\sigma}$  value uncertainties 
and ${\rm 1\sigma}$ upper limits, respectively. 
}}}
\end{picture}
} 
\label{fig1}
\end{figure} 
\begin{figure}[t]
\setlength{\unitlength}{1mm}
\resizebox{8.7cm}{!}{
\begin{picture}(80,76)(0,0)
\put ( -1.5,38) {\includegraphics[width=69.5mm, clip]{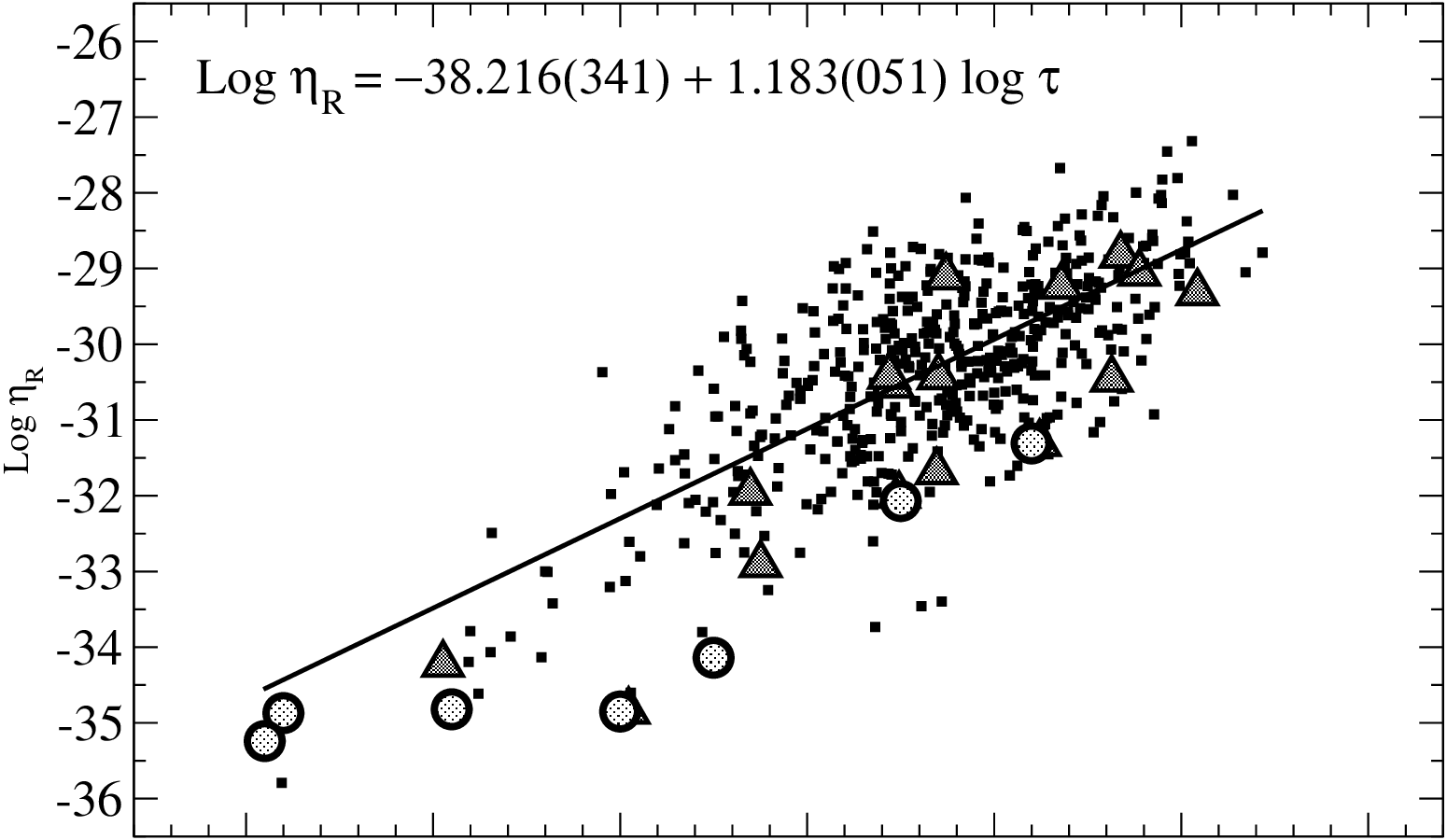}}
\put ( 0,-7) {\includegraphics[width=68mm, clip]{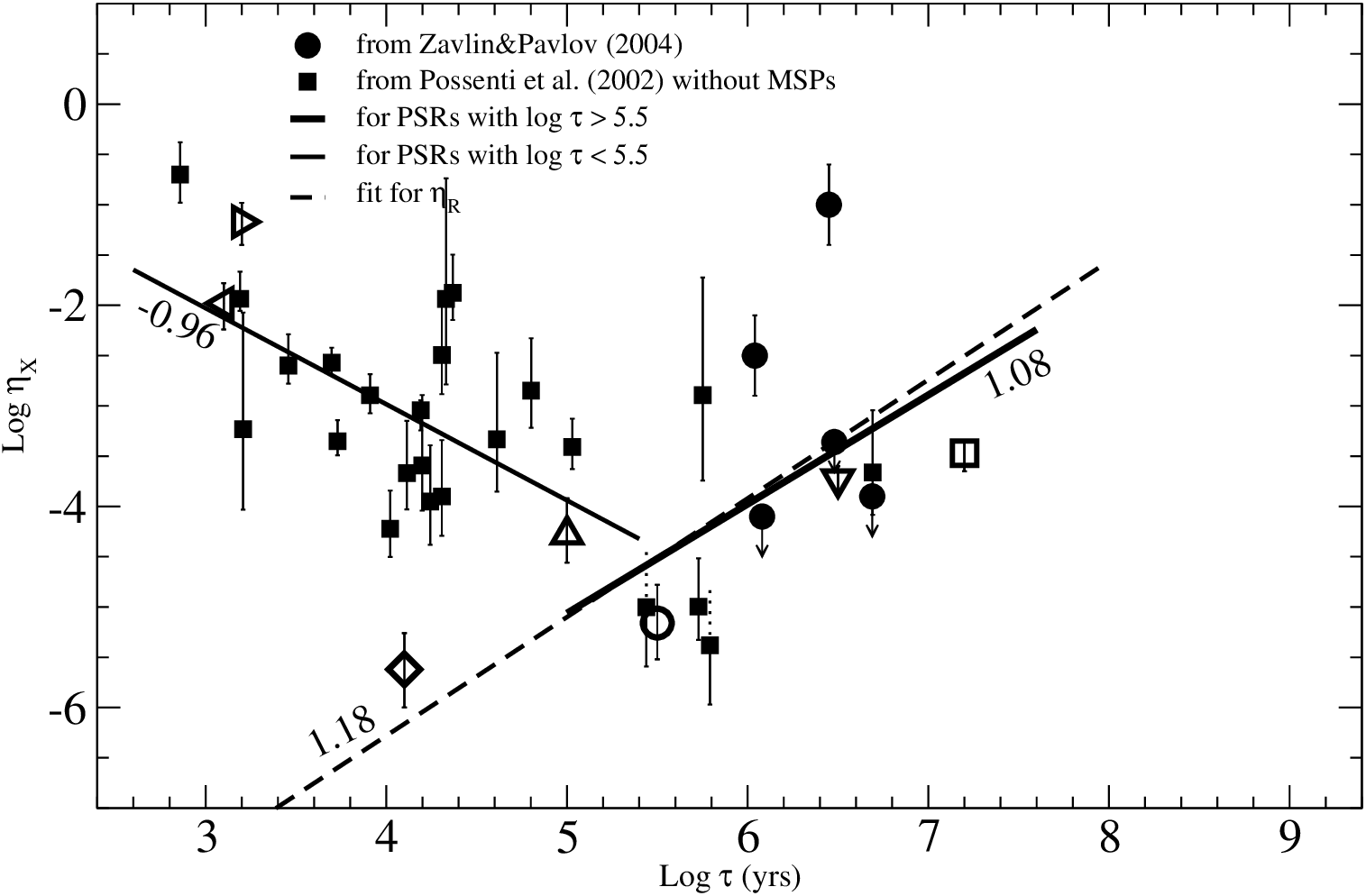}}
\end{picture}
} 
\caption{{\sl Top:} radio efficiencies of 500 pulsars from ATNF \cite{http} {\it vs} dynamical age; triangles mark the
pulsars with more accurate parallax-based distances and  circles
select optical pulsars from Table 1. {\sl Bottom:} X-ray efficiencies {\it vs} dynamical age; 
 lines are the linear regression fits of evolution tracks of younger and older pulsars and numbers show the line  slopes.}
\label{fig2}
\end{figure}  
\begin{figure}[tb]
\includegraphics[width=73mm,clip]{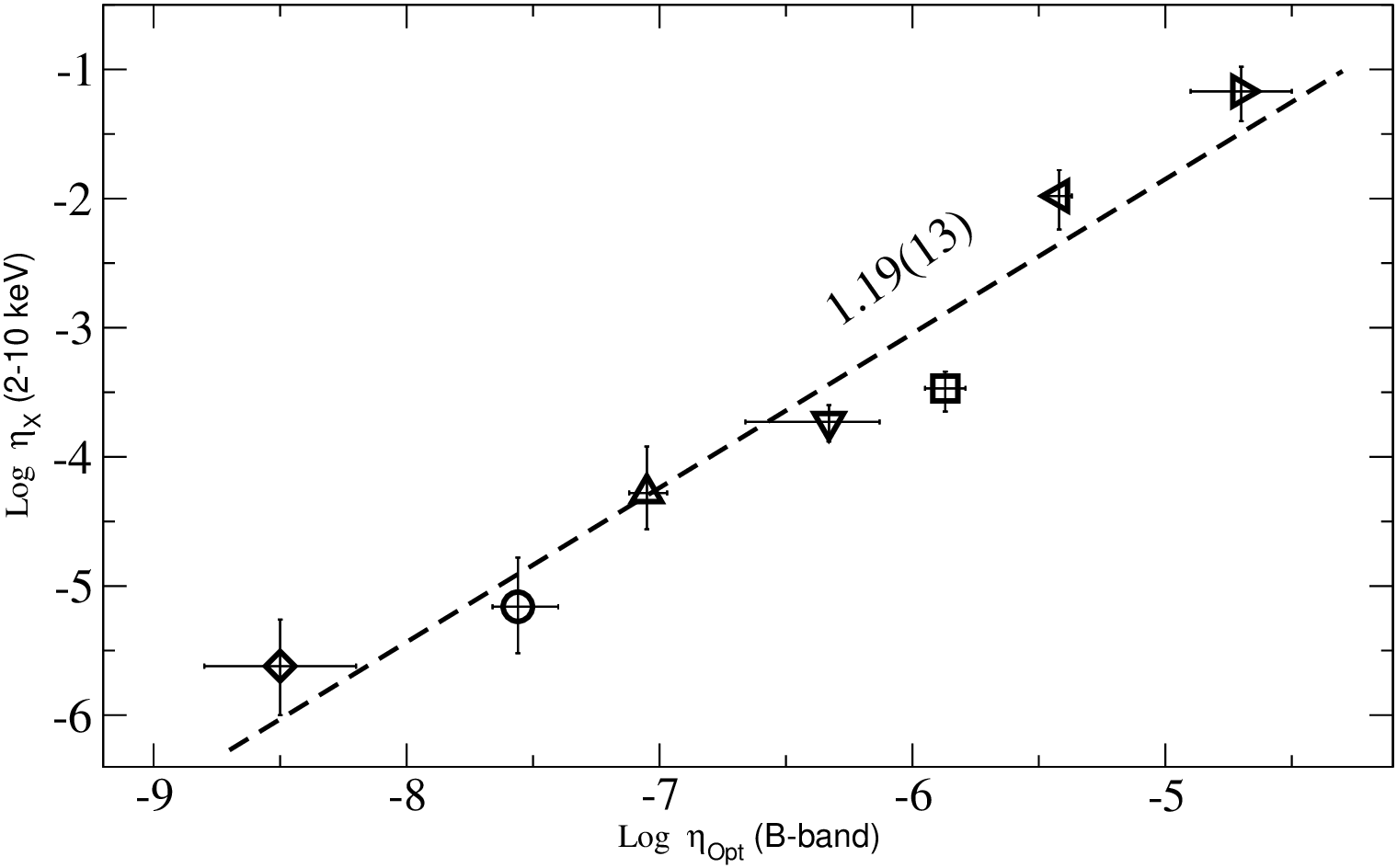}
\vspace{-1.1cm}
\caption{Relationship between the optical and X-ray efficiencies in the B band and 2-10~keV ranges.} 
\label{fig4}
\vspace{-0.7cm}
\end{figure} 
\begin{figure}[tb]
\includegraphics[width=73mm,clip]{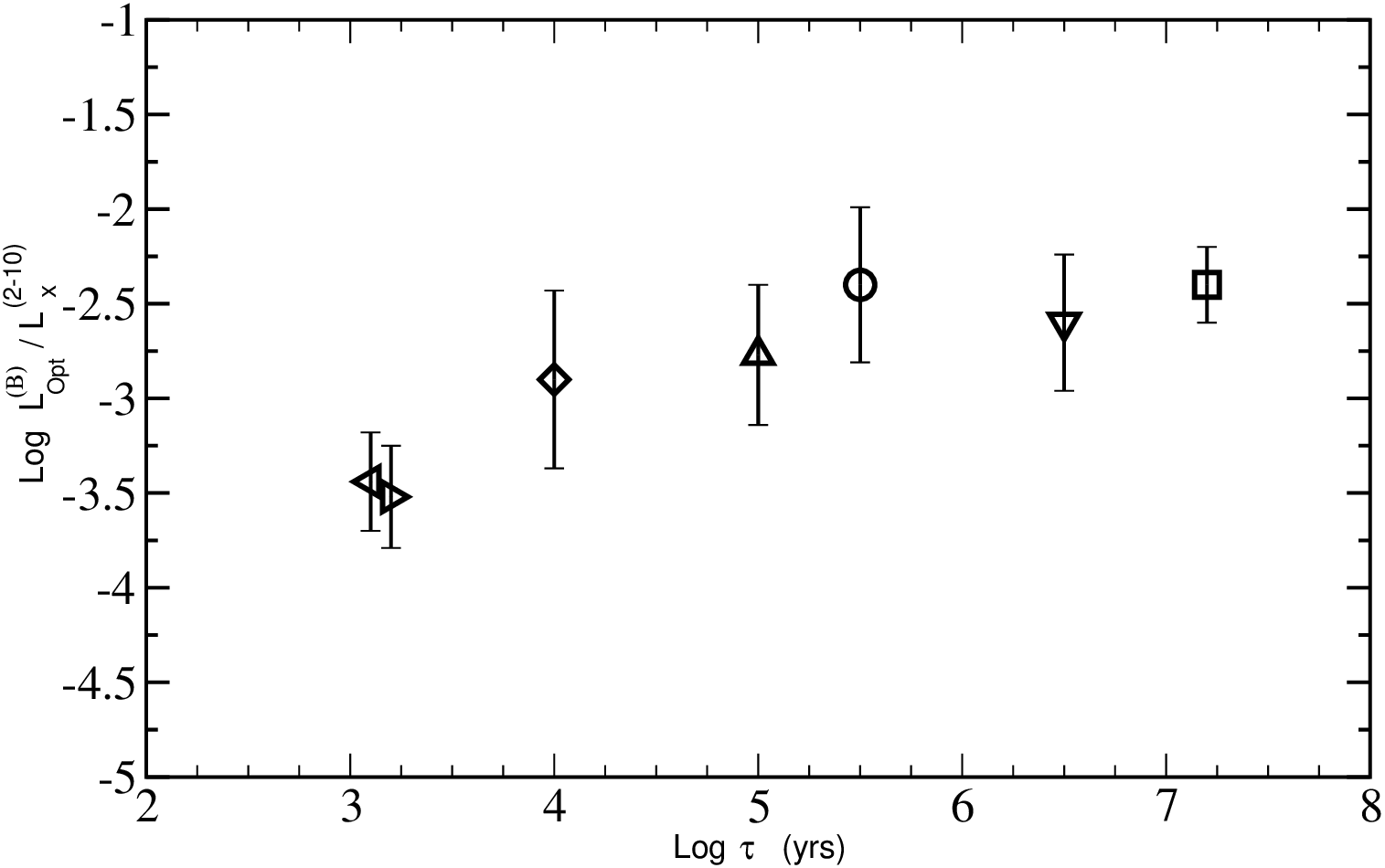}
\vspace{-1.1cm}
\caption{The evolution of the optical and X-ray luminosities ratio with dynamical  age.} 
\label{fig5}
\vspace{-.7cm}
\end{figure} 
%%%%%%%%%%%%%%%%%%%%%%%%%%%%%%%%%%%%%%%%%%%%%%%%%%%%%
\section{The   sample and data sources}
%%%%%%%%%%%%%%%%%%%%%%%%%%%%%%%%%%%%%%%%%%%%%%%%%%%%
We  have  analyzed multiwavelength  nonthermal  luminosities of  seven
pulsars detected  in the  optical range (see  Table 1; \cite{Zhar1,ZharS}).
 The  advantage  of  this sample  is  a  small  ($<$10\%)
uncertainty in the distances to the pulsars, determined from the radio
or optical  parallax measurements.  This strongly  minimizes errors in
the inferred luminosities.  Distances  for a few other optical pulsars
are much less certain and we excluded them from  consideration.
In the  optical we used the  data obtained in the  B-band \cite{Zhar1,ZharS}
 since practically all the pulsars from our sample were
detected  in this  band and  fluxes in  other bands  are  not strongly
different from these. Using this band  also allows us 
to  escape  possible contamination  of  the  pulsar  fluxes by  strong
nebular  lines in  cases when  an  associated nebula  around a  pulsar
exists, or is expected to exit.
In X-rays  we used the  data obtained in  2-10 keV range  \cite{Poss,ZharS}, where the nonthermal 
 power law spectral tails dominate over
the thermal  components detected from %middle-aged 
several pulsars  in a softer
energy range. 
%In some case these data were updated by new observations. 
 We added also the data from \cite{Zav} on 5
old pulsars recently observed with the Chandra and XMM.
In the radio and gamma-rays we used the data obtained at 400 MHz (ATNF
catalog \cite{http})  and in  the  EGRET range  of  400~MeV-10~GeV \cite{Thom},
respectively. Here we consider only isotropic equivalents of the luminosities neglecting 
the emission beaming  which is not yet properly known in the all considered ranges for all our objects.   
\vspace{-0.2cm}

%%%%%%%%%%%%%%%%%%%%%%%%%%%%%%%%
\section{Results}
% \label{}
%%%%%%%%%%%%%%%%%%%%%%%%%%%%
\subsection{Evolution}
%%%%%%%%%%%%%%%%%%%%%%%%%%%%%
The nonthermal luminosity and efficiency evolution of pulsars in different
spectral domains, based on the selected sample of the optical pulsars,
is shown in Fig.~1. We note the following points:\\
(1)		the spindown luminosity decreases monotonically with
		age, as expected from a formal dependence ${\rm \dot{E} \sim \tau^{-1} P^{-2}}$ and longer periods of older pulsars; 
(2)		the optical, X-ray and gamma-ray luminosities also
		decrease with age, but become almost constant starting
		from a middle-age epoch of $10^4- 10^5$ yr; 
(3)		the radio luminosity is less monotonic but starts to
		increase gradually approximately from the same age; 
(4)		as a result, the evolution of the optical and X-ray
		efficiencies is non-monotonic and shows a pronounced
		minimum efficiency at a middle-age epoch  of $10^4-10^7$ years and
		comparable higher efficiencies for younger and older
		pulsars in contrast to the conception of a monotonic decrease of the efficiency with age suggested by \cite{Goldoni} through lack of data on old pulsars at that time; 
(5)		the shapes of the dependencies $\eta_{opt}(\tau)$ and $\eta_X(\tau)$ 
                are remarkably similar to each other;  
(6)		the radio and gamma-ray efficiencies increase with age
		but their increases becomes apparently steeper from the
		same middle-age epoch.

 The fact that older pulsars are much more efficient radio emitters
 than younger ones is 
 %firmly 
 confirmed at a much higher significance
 level based on a rich sample of radio pulsars  extracted 
from the ATNF catalog (Fig.~2, top panel). %\ref{fig2}).  
The scatter of the data along the 
%general
 mean 
 evolution track shown in this panel   %\ref{fig2}
 by 
% stringht 
 solid 
 line may be explained by
 different geometrical/beaming factors and 
 %large 
 uncertainties in the
 distance. The  pulsars with 
parallax based distances are generally below this line.     
%the best fitting law.  
This may reflect a systematic shift between  the parallax and dispersion 
measure distances used  for the rest objects.

The  
% fact of 
% the
  nonmonotonic efficiency evolution in X-rays becomes  
much less
 %not 
 obvious if we consider a larger sample of pulsars identified  %mainly
 in X-rays including a few of them identified in the optical and X-rays (Fig.~2. low panel). %\ref{fig3}). 
 %Possible reasons are a higher uncertainty of the
 %X-ray data and smaller statistics than in radio range.  
 The statistics is much poor than in the radio and this  
 picture 
 %in Fig.\ref{fig3} 
 is 
% stronger
more affected by the distance and flux
 uncertainties. Nevertheless, applying  linear regression fits separately to sub-samples of sub-middle-aged and post-middle-aged pulsars, 
 one can resolve hints of the nonmonotonic
 evolution  clearly seen in the optical sub-sample.
In the example shown in Fig.2, (low panel) the Geminga age of $10^{5.5}$ yr, when the optical efficiency start to 
increase, was used to divide by sub-samples. In this case
 the evolution track
 of older pulsars may be compatible with what is seen in the radio range (cf. dashed and solid lines). 
Using other dividing ages from the range of the middle-aged epoch of $10^4-10^6$ yr does not change the picture qualitatively 
though slopes values can be different, but they are always negative and positive for younger and older pulsars, respectively.  
In all of the X-ray plots we added X-ray data for  5 
 old pulsars detected recently with Chandra and XMM \cite{Zav}. These data 
 are likely to be consistent  
% in agreement 
with the proposed evolution picture. 
 % and may even strengthen it. 
To confirm or reject it better quality X-ray data and  distance information 
are needed. %necessary.
%%%%%%%%%%%%%%%%%%%%%%%%%%%%%%%%%%%%%%%%%%%
%
%%%%%%%%%%%%%%%%%%%%%%%%%%%%%%%%%%%%%%%%%%%%%%%%%%%%%%%%%%%%%%%%%%%%
\subsection{Correlation between the optical and X-ray emission}
%%%%%%%%%%%%%%%%%%%%%%%%%%%%%%%%%%%%%%%%%%%%%%%%%%%%%%%%%%%%%%%%%%%%%
Similar shapes of the optical and X-ray evolution tracks suggest a
correlation between the luminosities in these domains. This is
confirmed by the linear regression fit of the $\eta_{opt}$ {\it vs} $\eta_X$ 
distribution shown in Fig.~\ref{fig4}. The correlation coefficient is
0.97. A strong correlation implies the same origin for the emission in
both spectral domains. It is independent of the
difference in the spectral slopes in these ranges observed for
the pulsars of different age. For instance,   
the extrapolation of the nonthermal component 
of the X-ray spectrum of young Crab-like
pulsars strongly overshoots the observed optical fluxes, 
while for older pulsars
 the optical fluxes are generally compatible with such 
extrapolation. This is reflected 
in the evolution of the ratio $L_{opt}/L_X$ shown in Fig.~\ref{fig5}.
 The ratio becomes almost
independent of age starting from a middle-age epoch.
%%%%%%%%%%%%%%%%%%%%%%%%%%%%%%%%%%%
 \begin{figure}[tb]
\includegraphics[width=64mm,clip]{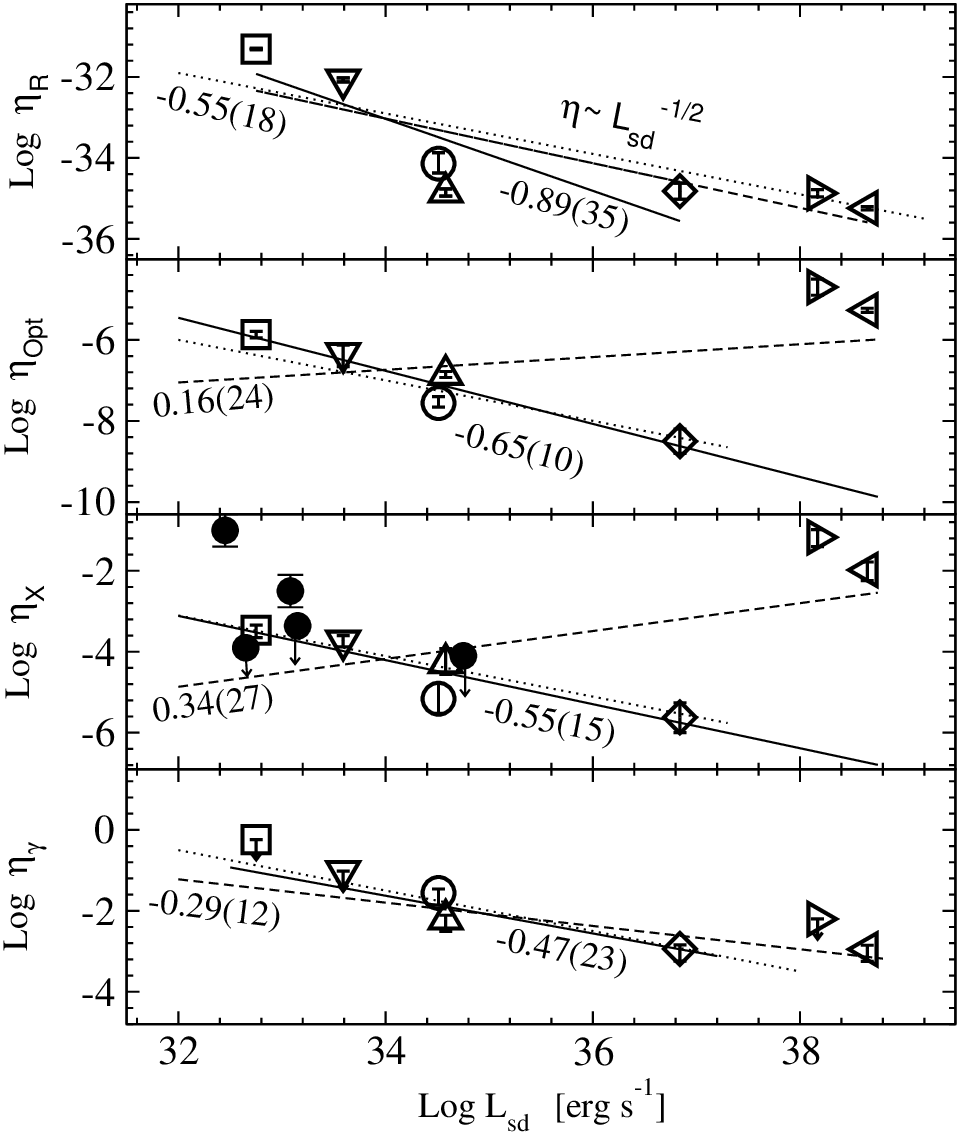}
\vspace{-1cm}
\caption{{\it From top to bottom:} relationships between the
efficiencies in the radio, optical, X-ray, and $\gamma$-ray spectral
domains and spindown luminosity.  Different pulsars are denoted by the
same symbols, as in Fig.~1. Dashed and solid lines show the best linear
regression fits of the whole set of pulsars, and without the two younger
ones (Crab and PSR B0540-69, with the highest $L_{sd}$),
respectively. The slope values are indicated near the lines.  Dotted
lines indicate a slope in case of the linear proportionality of the
luminosity to the Goldreich-Julian current \cite{Goldreich}.}
\label{fig6}
\vspace{-0.7cm}
\end{figure}
 \begin{figure}[tb]
\setlength{\unitlength}{1mm}
\resizebox{8.7cm}{!}{
\begin{picture}(80,77)(0,0)
\put ( -1,36) {\includegraphics[width=69mm, clip]{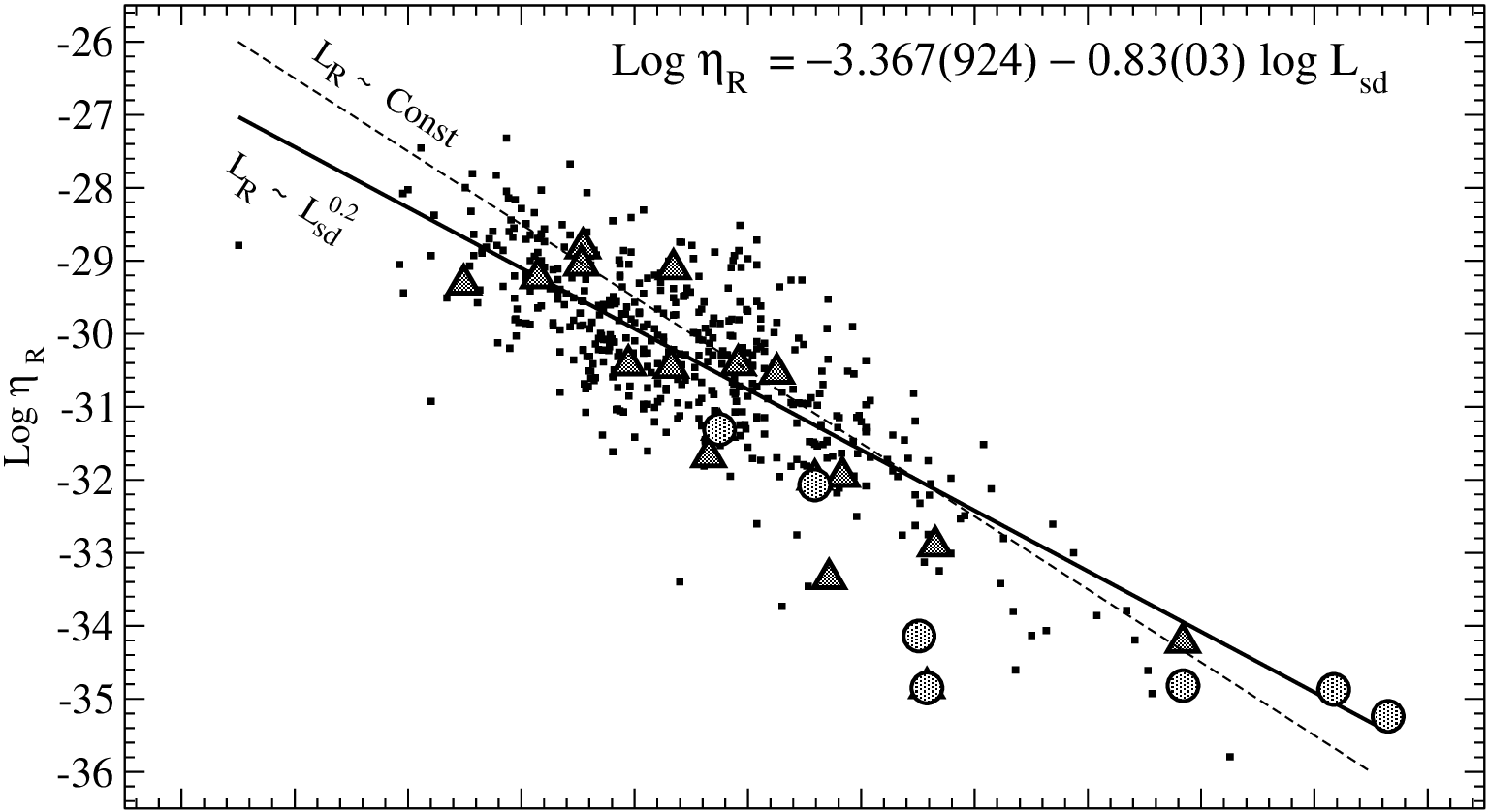}}
\put ( 0,-7) {\includegraphics[width=68mm, clip]{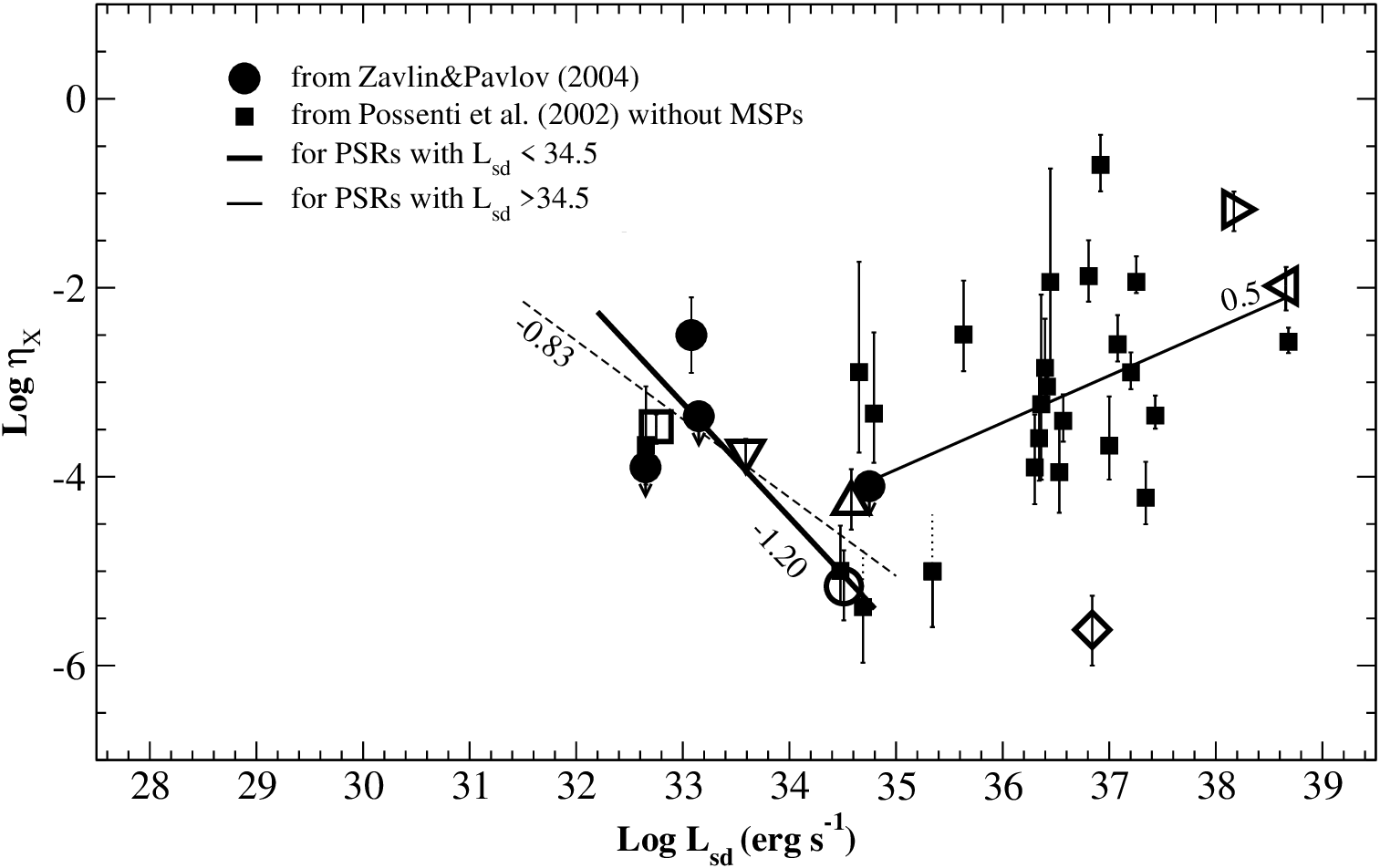}}
\end{picture}
} 
%\includegraphics[width=66mm,clip]{zharikovfig7.eps}
%\vspace{-1cm}
\caption{ {\sl Top:} radio efficiencies of 500 pulsars from ATNF  database \cite{http} {\it vs} $L_{sd}$; 
notations are the same as in Fig.~2. 
%triangles mark the
%pulsars with more accurate parallax-based distances and  circles
%select optical pulsars from Table 1. 
{\sl Bottom:} X-ray efficiencies  {\it vs}  spin-down luminosity;  
lines fit the distribution of lower and higher $L_{sd}$ pulsars;
numbers show the line slopes.}
\label{fig7}
\vspace{-0.6cm}
\end{figure}
%
% \begin{figure}[tb]
%\includegraphics[width=66mm, clip]{zharikovfig8.eps}
%\vspace{-1cm}
%\caption{ The X-ray efficiencies  {\it vs}  spin-down luminosity. Lines fit the distribution of lower and higher $L_{sd}$ pulsars;
%numbers show the line slopes.}
%\label{fig8}
%\vspace{-0.8cm}
%\end{figure}
%%%%%%%%%%%%%%%%%%%%%%%%%%%%%%%%%%%%%%%%%%%%%%%%%%%%%%%%%%
\subsection{The efficiency and the spindown luminosity}
%%%%%%%%%%%%%%%%%%%%%%%%%%%%%%%%%%%%%%%%%%%%%%%%%%%%%%%%%%%%%%%%
The distributions of the efficiencies of the optical pulsars in different
spectral domains {\it vs} spindown luminosity are shown in Fig.~\ref{fig6}.  These
distributions demonstrate that: 
(1)		in the radio and gamma-rays, less energetic pulsars
		(with low $L_{sd}$) are more efficient photon emitters than
		more energetic ones; 
(2)		in these ranges the efficiency increases monotonically
		with the $L_{sd}$ decrease;  
(3)		the same is seen in the optical and X-ray ranges with
		an exclusion of the two most energetic and young
		pulsars, Crab and PSR B0540-69, which form a separate
		sub-class in these distributions; 
(4)		recent data for the five old pulsars identified in X-rays
		(filled circles  in Fig.~5) do not change the above
		tendency in X-rays and may even strengthen it (cf.~Fig.~\ref{fig7}, bottom panel).

As in case of the evolution picture, the fact that the pulsars with
lower $L_{sd}$ are more effective radio emitters is firmly confirmed by the
analysis of a much larger sample of pulsars extracted from the ATNF
catalog \cite{http} (Fig.~\ref{fig7}, top panel). 
Comparing the triangles and black point distributions in Fig.~\ref{fig7}, (top panel),
 we can
note that the scatter around an average distribution shown by  the solid
line is likely equally caused by uncertainties in the pulsar
geometry (triangle points), and DM-based distances (black points).
 The slope of the distribution is different from a trivial one assuming  
$L_R\sim {\it const}$ (dashed line in this plot) but suggests an increase of 
the radio luminosity  $\propto  L_{sd}^{0.2}$. 
     
The separation into two sub-classes of the low and high spindown luminosity
pulsars that is  clearly seen in the middle panels of Fig.~\ref{fig6} for %the sample of
the optical pulsars is not  obvious for a wider sample of pulsars
identified  in X-rays (Fig.~\ref{fig7}, bottom panel). The reasons are likely to be the same  
as we have mentioned in Sect.~3.2 (poor statistics + distance and flux uncertainties).  
%likely in a poor
%statistics, particularly for a low spindown subsample of this
%distribution, and a large uncertainty in distances, which play a
%higher role here in comparison with the radio range where the pulsar
%statistics is much more representative. 
 However, as in the case of the evolution picture for the
 same sample shown in Fig.~\ref{fig4}, 
%applying the 
linear regression fits allow us %it is possible 
to resolve a hint of a bimodal distribution with the efficiency increases toward the low and high
%edges
sides 
of the considered $L_{sd}$ range. The bottom panel of Fig.~\ref{fig7}, where $L_{sd}$ of Geminga was used, 
as in case of Fig.~\ref{fig4}, to divide the pulsars by sub-samples,  suggests also that the
distribution of the low $L_{sd}$ pulsars  may be roughly compatible
with what is seen in the radio range (cf. solid and dashed lines in this plot). % Fig~\ref{fig7}).
The suggested bimodal distribution can hardly be explained by geometry effects, if
we compare the relatively small  scattering of the triangle points
 around a mean distribution  
in the radio 
%Fig.~\ref{fig7}, 
(top panel), likely caused by geometry effects, 
and the much larger scattering of all points in X-rays 
%Fig.~\ref{fig7}, 
(bottom panel). 
%
%%%%%%%%%%%%%%%%%%%%%%%%%%
\section{Summary}
%%%%%%%%%%%%%%%%%%%%%%%%%%%%%
	Our multiwavelength analysis of the nonthermal luminosity and
	efficiency distributions {\it vs} dynamical age and spindown power
	of the pulsars detected in the optical range suggests significantly  
	nonmonotonic evolution of their emission in the optical and 
	X-ray domains and/or a bimodal distribution of the respective
	efficiencies {\it vs} spindown power. 
Less energetic old pulsars are much more efficient photon
 emitters than the middle-aged ones and their efficiency is comparable
 with that for the much younger and energetic Crab-like objects.
%The relatively
%	low efficiency level of middle-aged pulsars increases for the 
%       less energetic
%	old pulsars which are much more efficient photon emitters as are young
%	and energetic Crab-like objects. %
%%
%
The bimodal structure is  not 
	significant but still resolved in a
% wider 
larger
sample of pulsars
	identified  in X-rays, where it may be 
smoothed
%dimed out 
%In this sample 
	%It may be hidden  in 
	by large distance and flux uncertainties and needs further study.  % in the distances to the
	%pulsars and in their fluxes. % values. 
	%and needs further studies. 
	An apparent evolution track of old
	pulsars in the optical and X-rays is roughly compatible with the
	track seen in the radio range where the evolution is
	accompanied by a significant increase of the radio-efficiency
	with NS  age. If it is true, this implies %suggests 
	an universal engine that drives the
	non-thermal evolution of old pulsars by a unified way
	increasing their efficiency with age in the all spectral domains.\\

 This work has been partially supported by CONACYT project 36585-E, 
 and RFBR grants 02-02-17668, 03-02-17423, 03-07-90200 and RLSS 1115.2003.2

\end{document}